\documentclass{pj}
\usepackage[pdftex]{graphicx}
\usepackage{amssymb}
\usepackage{hyperref}[2003/11/30]
\usepackage{color}
\usepackage{verbatim}
\usepackage{amsmath}
\usepackage{upgreek}
\usepackage{epstopdf}
\usepackage{cite}

\begin{document}
\setcounter{page}{1}
\pjheader{Vol.\ x, y--z, 2004}

\title{Permittivity of waste-activated sludge by an open-ended coaxial line}

%

\author{J.~S.~Bobowski, T.~Johnson, and C.~Eskicioglu}

\address{School of Engineering, University of
British Columbia Okanagan\\
3333 University Way, Kelowna, British
Columbia V1V 1V7, Canada}

\runningauthor{Bobowski}
\tocauthor{J.~Bobowski}

\begin{abstract}
The complex permittivity of thickened waste activated sludge (WAS) was measured from 3~MHz to 40~GHz.  The solid content of the thickened WAS sample was varied from 4.5\% to 18\% by weight. The permittivity spectra exhibit features typical of biological tissues that have a high water content.  At high frequencies, a Debye-type dispersion is observed with a relaxation rate of 19 GHz characteristic of the bulk water in the sample ($\gamma$-dispersion).  At lower frequencies, the solid content of the samples determines the properties of the permittivity.  The onset of the so-called $\beta$-dispersion, attributed to the charging of cell membranes, occurs between 10-100~MHz.  For samples with higher solid concentrations, a weak dispersion of the real part of the permittivity, characteristic of bound water, was observed at intermediate frequencies ($\delta$-dispersion).
\end{abstract}

\section{Introduction}
\label{sec:intro}

As a byproduct of domestic (sewage) and industrial wastewater treatment plants, waste activated sludge (WAS) is pervasive.  WAS is comprised of different groups of microorganisms, organic and inorganic matter agglomerated together in a polymeric network formed by microbial extracellular polymeric substances (EPS) and cations~\cite{Li:1990}.  Treating and then disposing of WAS is costly and can negatively impact the environment. Anaerobic digestion, in which microorganisms are used to breakdown organic matter in the absence of oxygen, is commonly used to stabilize and reduce the volume of the WAS.  In the final stage of anaerobic digestion, a methane-rich biogas is produced which can be collected and later used as fuel.  The time it takes for anaerobic digestion to reach completion can be as high as 20~days and can be a bottleneck in the treatment of wastewater.  For a comprehensive review of anaerobic digestion of WAS see Ref~\cite{Appels:2008}.

Efforts to pretreat WAS before anaerobic digestion to enhance the rate of methane production have shown promise.  The pretreatment methods include thermal cycling, sonication, application of high pressure, chemical and enzymatic treatment, microwave irradiation, and exposure to high-intensity pulsed electric fields~\cite{Appels:2008, Eskicioglu:2009, Saha:2011, Salerno:2009, Lee:2010, Sheng:2011}.  The goal of all pretreatment methods is to enhance the bioavailability of the WAS to the methane-forming microorganisms in the anaerobic digester.  These enhancements can be achieved by puncturing or otherwise damaging cell membranes and/or cell walls, breaking down or reducing the characteristic size of the floc structure, and converting complex organic molecules to simpler forms~\cite{Salerno:2009}.

To this point, the microwave treatments have been restricted to frequencies of 2.54~GHz and 915~MHz due to the availability of inexpensive high-power microwave tubes (magnetrons) used in conventional and industrial microwave ovens~\cite{Saha:2011,Park:2004,Eskicioglu:2007,Hong:2006}.  However, there are no fundamental reasons to suppose that these are the optimal frequencies with which to pretreat WAS.  Additionally, the method of irradiating the samples must be carefully evaluated.  A load impedance presented by the sample that is poorly matched to the source impedance of the generator will result in an inefficient system in which not all of the incident power is absorbed by the sample.  The objective of this work was to characterize the complex permittivity of WAS over a broad frequency range and to identify the mechanisms leading to the observed dispersions.  The measurements and theoretical insights into the active microscopic mechanisms allow for the selection of candidate frequencies with which to optimize the electromagnetic pretreatment of WAS.  Furthermore, custom applicators can be designed and constructed such that the incident signal power is very efficiently transferred to the sample.

\section{Experimental Methods}
\label{sec:methods}

When a material is exposed to a time-harmonic electric field $\mathbf{E}e^{j\omega t}$ of angular frequency $\omega$, the total current density $\mathbf{J}$ is the sum of the conduction and displacement current densities:
\begin{equation}
\mathbf{J}=\left(\sigma_\mathrm{dc}+j\omega\varepsilon_0\varepsilon_\mathrm{r}\right)\mathbf{E}=\left[\left(\sigma_\mathrm{dc}+\omega\varepsilon_0\varepsilon^{\prime\!\prime}\right)+j\omega\varepsilon_0\varepsilon^\prime\right]\mathbf{E},\label{eq:J}
\end{equation}
where $\varepsilon_0$ is the permittivity of free space and $\sigma_\mathrm{dc}$ and \mbox{$\varepsilon_\mathrm{r}=\varepsilon^\prime-j\varepsilon^{\prime\!\prime}$} are the dc conductivity and relative permittivity of the material.  In this work, the electrical properties ($\varepsilon^\prime$, $\varepsilon^{\prime\!\prime}$, and $\sigma_\mathrm{dc}$) of WAS were measured using open-ended coaxial probes.  The open-ended coaxial probe was chosen because it is a broadband ($10^5$--$10^{10}$~Hz) measurement suitable for liquid or semisolid samples and requires no sample preparation.  There exist numerous other measurement techniques.  Parallel plate geometries are useful for frequencies below 30~MHz.  Transmission measurements are broadband but, in their simplest configuration, require samples machined to fit tightly inside a waveguide or coaxial transmission line.  Resonant cavities are very precise, but operate only at discrete frequencies~\cite{Kaatze:2006}.

\begin{figure}[t]
\centerline{\includegraphics[width=0.7\columnwidth]{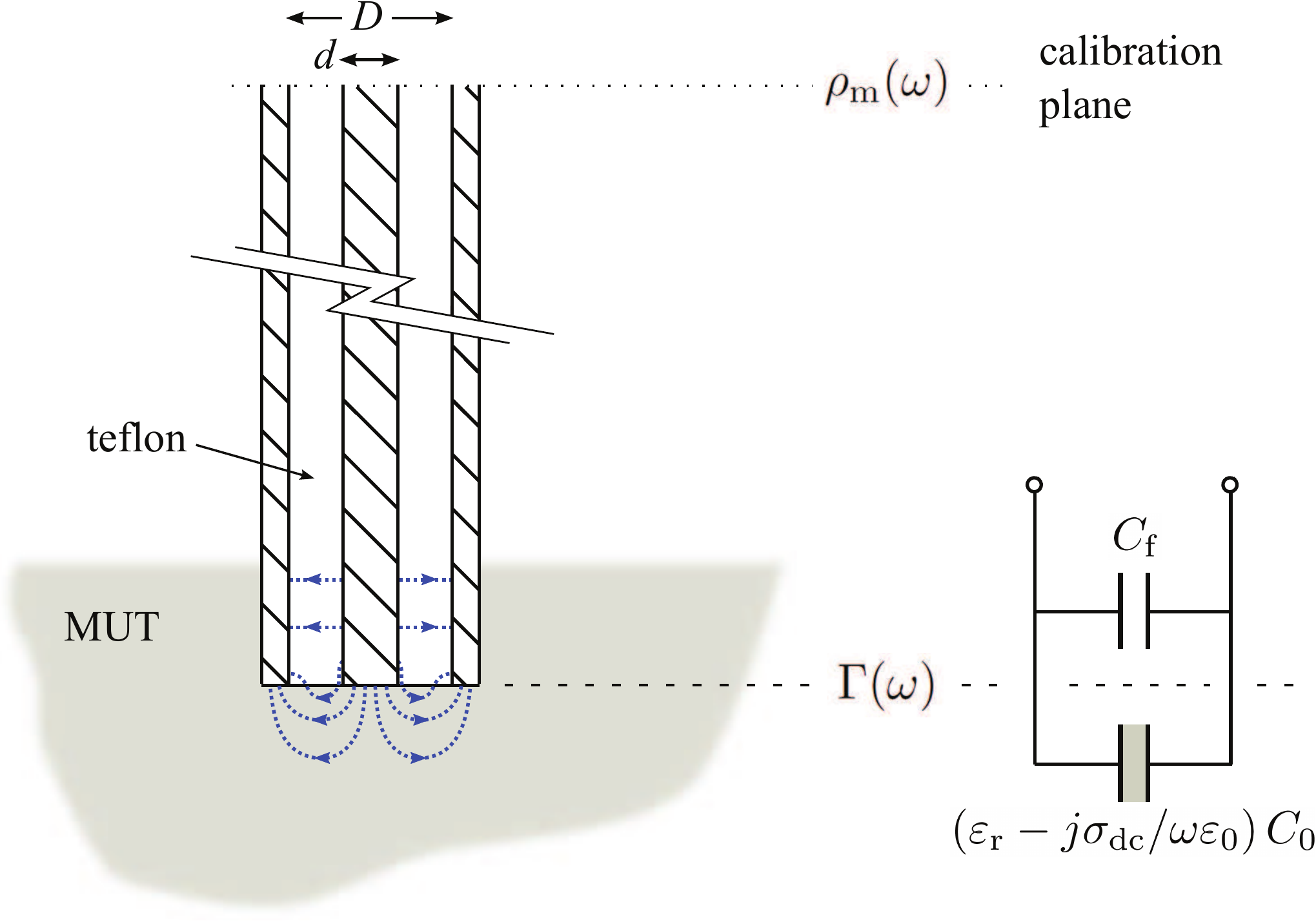}}
\caption{A schematic of the experimental geometry showing the open end of the coaxial probe submerged in the MUT.  The calibration plane of the VNA is established at the opposite end of the probe.  The reflection coefficient $\rho_\mathrm{m}(\omega)$ is measured directly by the VNA and can be related to the reflection coefficient $\Gamma(\omega)$ at the probe tip.  The dotted lines represent electric field lines in the vicinity of the interface between the probe tip and the MUT.  The impedance of the interface is modeled as parallel shunt capacitances $C_\mathrm{f}$ and $\left(\varepsilon_\mathrm{r}-j\sigma_\mathrm{dc}/\omega\varepsilon_0\right)C_0$.}
\label{fig:fringe}
\end{figure}

The probes were fabricated from short ($\approx 20$~cm) sections of commercially available semirigid coaxial transmission line.  For frequencies less than 1~GHz, UT-141 coaxial cable was used (shield inside diameter \mbox{$D=2.99$~mm}, center conductor diameter \mbox{$d=0.92$~mm}).  At higher frequencies, evanescent modes at the probe tip can be excited resulting in a breakdown of the measurement technique.  These unwanted modes can be suppressed by switching to a smaller diameter probe.  Measurements up to 40~GHz were made using UT-085 coaxial cable (\mbox{$D=1.67$~mm}, \mbox{$d=0.51$~mm}).  One end of the probe was fitted with a connector and the opposite end was polished flat creating an open circuit condition.  During the measurements, the open end of the probe was submerged into the material under test (MUT) resulting in a mismatched load.  Signals incident from a vector network analyzer (VNA) are partially reflected at the MUT-probe interface.  By analyzing the amplitude and phase of the reflection coefficient as a function of frequency, the complex permittivity and dc conductivity of the MUT were determined.  The experimental geometry is shown in Fig.~\ref{fig:fringe}.

The reflection coefficient $\Gamma$ at the probe tip is determined from the characteristic impedance $Z_0=50~\Omega$ of the coaxial transmission line and the effective load impedance $Z_\mathrm{L}$ present at the interface between the tip and the MUT:
\begin{equation}
\Gamma(\omega)=\frac{Z_\mathrm{L}-Z_0}{Z_\mathrm{L}+Z_0}.\label{eq:Gamma}
\end{equation}
This load impedance is typically modeled as two parallel shunt capacitors $C_\mathrm{f}$ and $C(\varepsilon_\mathrm{r},\sigma_\mathrm{dc})$. Here, $C_\mathrm{f}$ is purely reactive and accounts for fringing of the electromagnetic fields that occurs within the dielectric of the coaxial probe whereas $C(\varepsilon_\mathrm{r},\sigma_\mathrm{dc})=\left(\varepsilon_\mathrm{r}-j\sigma_\mathrm{dc}/\omega\varepsilon_0\right)C_0$ accounts for fringing that occurs within the MUT~\cite{Stuchly:1980}.  The load admittance at the probe tip can then be written:
\begin{equation}
Z_\mathrm{L}^{-1}=C_0\left(\omega\varepsilon^{\prime\!\prime}+\frac{\sigma_\mathrm{dc}}{\varepsilon_0}\right)+j\omega\left(C_\mathrm{f}+\varepsilon^\prime C_0\right).\label{eq:ZL}
\end{equation}
The measured reflection coefficient $\rho_\mathrm{m}$ at the calibration plane of the VNA can be related to $\Gamma$ by modeling the coaxial probe as a two-port network.  The relevant scattering parameters, which account for losses along the length of the probe and for spurious reflections, were determined using three calibration terminations: short-circuit, open-circuit, and a standard liquid of known permittivity~\cite{Kraszewski:1983}.  A measurement of the real and imaginary components of $\Gamma$ allows Eqns.~\ref{eq:Gamma} and \ref{eq:ZL} to be used to determine both $\varepsilon^\prime$ and $\varepsilon^{\prime\!\prime}+\sigma_\mathrm{dc}/\omega\varepsilon_0$ at each measurement frequency.

Prior to characterizing the electrical properties of WAS, the experimental technique was verified using two control samples.  In both cases, distilled water at \mbox{$25.0\pm0.1^\circ$C} was used as the standard liquid termination.  The first control sample was 99.9\% methyl alcohol which has negligible $\sigma_\mathrm{dc}$.  The relative permittivity extracted from $\Gamma$ was in excellent quantitative agreement with earlier measurements made by Bao and coworkers~\cite{Bao:1996}.  Our measurements of methyl alcohol extended from 10~MHz to 40~GHz.  The lower end of the experimental bandwidth is set by the resolution of the reflection coefficient measurement while the upper end is set by the onset of higher-order evanescent modes at the probe-MUT junction.

The second control sample was a 0.03~molarity solution of 99.8\% NaCl dissolved in distilled water.  The real and imaginary parts of \mbox{$\varepsilon_\mathrm{r}-j\sigma_\mathrm{dc}/\omega\varepsilon_0$} of the solution were extracted from the reflection coefficient measurements.  As expected, at high frequencies we achieved an accurate measurement of $\varepsilon_\mathrm{r}$ of pure water~\cite{Buchner:1999}.  At sufficiently low frequencies, $\varepsilon^{\prime\!\prime}$ is dominated by the $\sigma_\mathrm{dc}$ term such that the imaginary component of \mbox{$\varepsilon_\mathrm{r}-j\sigma_\mathrm{dc}/\omega\varepsilon_0$} varies as the inverse of frequency.  The intrinsic behavior of the real component, on the other hand, is independent of $\sigma_\mathrm{dc}$.  However, when submerged in an electrolyte, the coaxial probe acquires a surface charge resulting in a so-called electric double layer at the probe-MUT interface such that its effective impedance is modified.  This electrode polarization effect can be reliably modeled and corrected for by including a complex polarization impedance $Z_\mathrm{p}$ in series with the parallel combination of $C_\mathrm{f}$ and $C(\varepsilon_\mathrm{r},\sigma_\mathrm{dc})$ shown in Fig.~\ref{fig:fringe}~\cite{Schwan:1992, Raicu:1998}.  In our measurements of the NaCl solution, the effects of electrode polarization were clearly noticeable below 10~MHz.  The correction scheme reliably removed these effects allowing us to extend the bottom end of our measurement bandwidth by more than an order of magnitude.

\section{Waste-Activated Sludge Permittivity}\label{sec:WAS}

The open-ended coaxial probe was next used to determine the unknown permittivity of thickened WAS (TWAS) obtained from the Wastewater Treatment Facility (WWTF) in Kelowna, British Columbia, Canada.  As routine practice, before final disposal, TWAS samples are further dewatered at the WWTF using a centrifuge such that the solid content of the resulting material (called ``sludge cake'') is $18\pm 2\%$ by weight.  In this study, the permittivity and conductivity of both TWAS ($4.5\%$ solids) and sludge cake (18\% solids) were measured.  First, $\varepsilon_\mathrm{r}$ and $\sigma_\mathrm{dc}$ of the 18\% sample were determined from reflection coefficient measurements made using open-ended coaxial probes.  Next, distilled water was added to the 18\% sludge cake to produce a sample representative of TWAS, with a solid content of $4.5\pm0.5$\%, and the measurements of the electrical properties were repeated. We chose to dilute the sludge cake with distilled water so that we could directly compare data obtained from 18\%- and 4.5\%-solid solutions prepared from a single sample.  To be sure that the 4.5\% sample was a good representation of TWAS, we also measured a 4.5\% TWAS sample obtained directly from the Kelowna WWTF.  For all measurements, a 0.03~molarity NaCl solution was used as the liquid calibration standard as it mimics some of the essential electrical properties expected from the WAS samples.

\begin{figure}[t]
\centerline{\includegraphics[width=0.7\columnwidth]{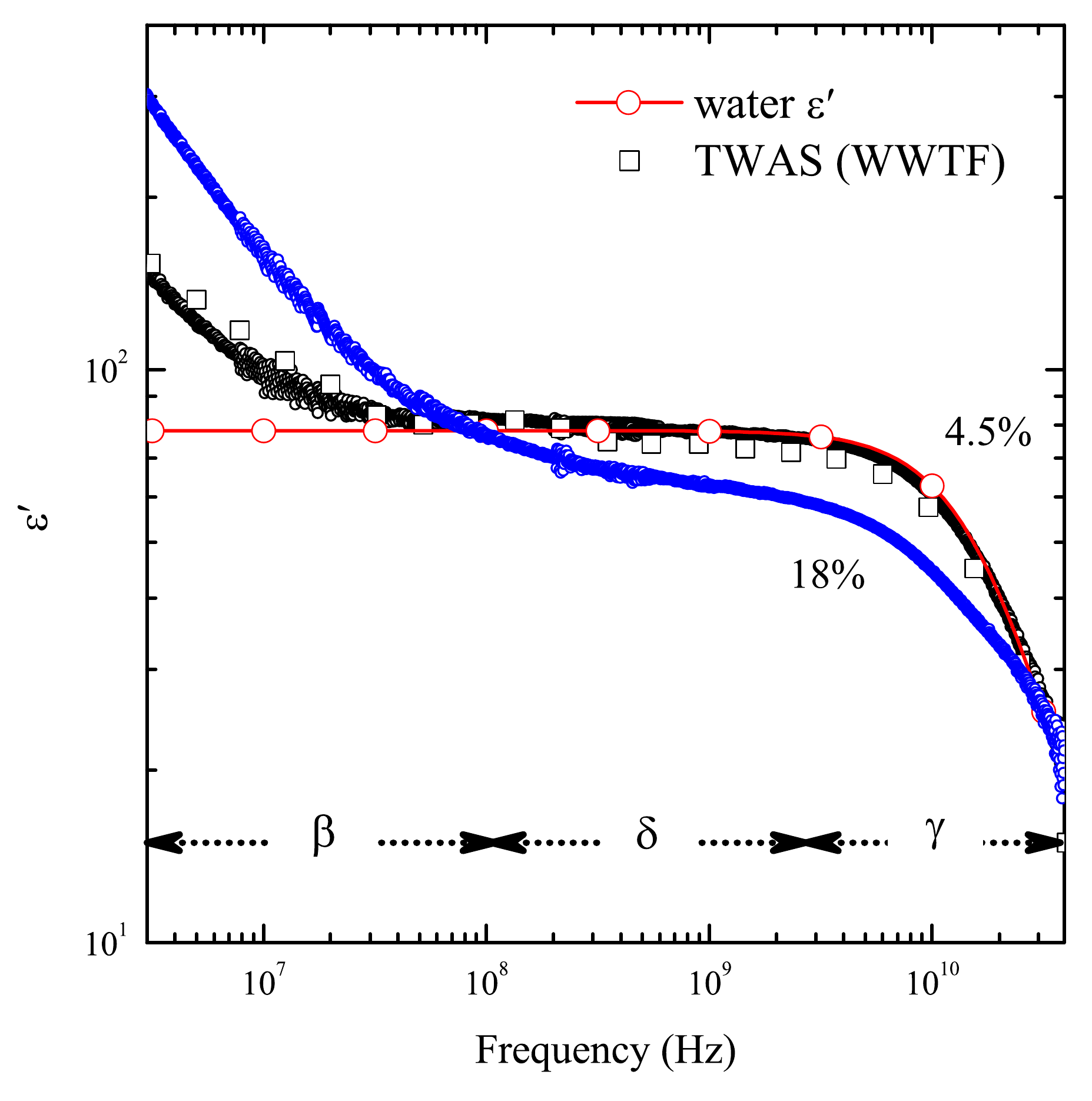}}
\caption{Plot of $\varepsilon^\prime$ of WAS as a function of frequency at 25$^\circ$C.  Samples with both 4.5\% (black) and 18\% (blue) solid concentrations were measured.  The electrode polarization artifacts have been removed and the low-frequency upturns are features intrinsic to the samples.  At 4.5\%, results for both sludge cake diluted with distilled water (black points) and TWAS from the WWTF (open squares) are shown. The solid red curve is the real component of the permittivity of pure water.}
\label{fig:WASpermRe}
\end{figure}

When interpreting the results of the permittivity and conductivity of the WAS samples, comparisons were made to the results of previous studies of various biological tissues.  For a thorough review see Ref.~\cite{Pethig:1984} and references therein.  Figure~\ref{fig:WASpermRe} shows the real component of \mbox{$\varepsilon_\mathrm{r}-j\sigma_\mathrm{dc}/\omega\varepsilon_0$} for both the 4.5\% and 18\% TWAS samples after correcting for electrode polarization effects.
For all TWAS samples, the reflection coefficient measurements were repeated at least five times.  In between measurements the probe tip was removed from the sample, cleaned, and then reinserted into the sample at a new location.  The reflection coefficients were very insensitive to the probe location indicating that the bulk material properties of the samples were measured.   Moreover, multiple TWAS samples were measured at both the 4.5\% and 18\% concentrations.  Consistent results were obtained among samples of the same concentration.  Note that $\varepsilon^\prime$ of the 4.5\% TWAS sample obtained from the WWTF is very similar to that of the sludge cake sample that has been diluted with distilled water.

Above 100~MHz, $\varepsilon^\prime$ of the 4.5\% sample is nearly identical to that of pure water.  This characteristic, referred to as $\gamma$-dispersion, is very common in biological samples and is due to the relaxation of polar water molecules.  Below 100~MHz, $\varepsilon_\mathrm{r}$ of the 4.5\% sample shows an upturn marking the onset of a distinct dispersion mechanism.  This effect, termed $\beta$-dispersion, is ubiquitous in samples of biological tissue and is typically active from 100~kHz to 10~MHz~\cite{Pethig:1984}.  The dispersion arises from the Maxwell-Wagner effect in which charge accumulates at the junction between materials of different permittivities and conductivities.  In biological samples, cell membranes form barriers between the intra- and extracellular fluids that in general have distinct electrical properties. The 18\% sample has a higher concentration of suspended cells and, as expected, the extracted $\varepsilon^\prime$ shown in Fig.~\ref{fig:WASpermRe} exhibits an enhanced $\beta$-dispersion below 100~MHz.

It is worth noting that $\beta$-dispersion may be a useful quantitative probe of cell viability in  WAS samples.  If a WAS sample is treated by a process that induces lysis, the number of intact cell membranes dividing the intra- and extracellular fluids would be reduced; presumably resulting in a suppression of $\beta$-dispersion.  It may be possible to use permittivity measurements in the range from 0.1 to 100~MHz as a quick and accurate initial indicator of treatment effectiveness.

In contrast to the 4.5\% sample, $\varepsilon^\prime$ of the 18\% sample shows no plateau marking the transition from one dispersion mechanism to the next.  Rather, the broad dispersive feature from 100~MHz to 10~GHz indicates that (i) the $\beta$-dispersion associated the charging of cell membranes extends to higher frequencies and (ii) there are other dispersion mechanisms active in this frequency range.  These features, generally termed $\delta$-dispersion, are attributed to a variety of mechanisms.  Most commonly $\delta$-dispersion is thought to arise from dielectric relaxation of water molecules bound to adjacent proteins.  The rotational relaxation time of bound water molecules is longer than that of bulk water causing the dielectric dispersion to be shifted to lower frequency.  Much of the bound water will be associated with proteins present in cell membranes~\cite{Pethig:1984}.  It is plausible that exposing WAS to electromagnetic radiation in the UHF range can bring about physiological changes that may enhance anaerobic digestion.  For example, if tuned to the relaxation rate of water molecules bound to membrane proteins, the electromagnetic energy will be, in part, absorbed directly by the intracellular fluid which may lead to cell lysis after sufficient exposure.

\begin{figure}[t]
\centerline{\includegraphics[width=0.7\columnwidth]{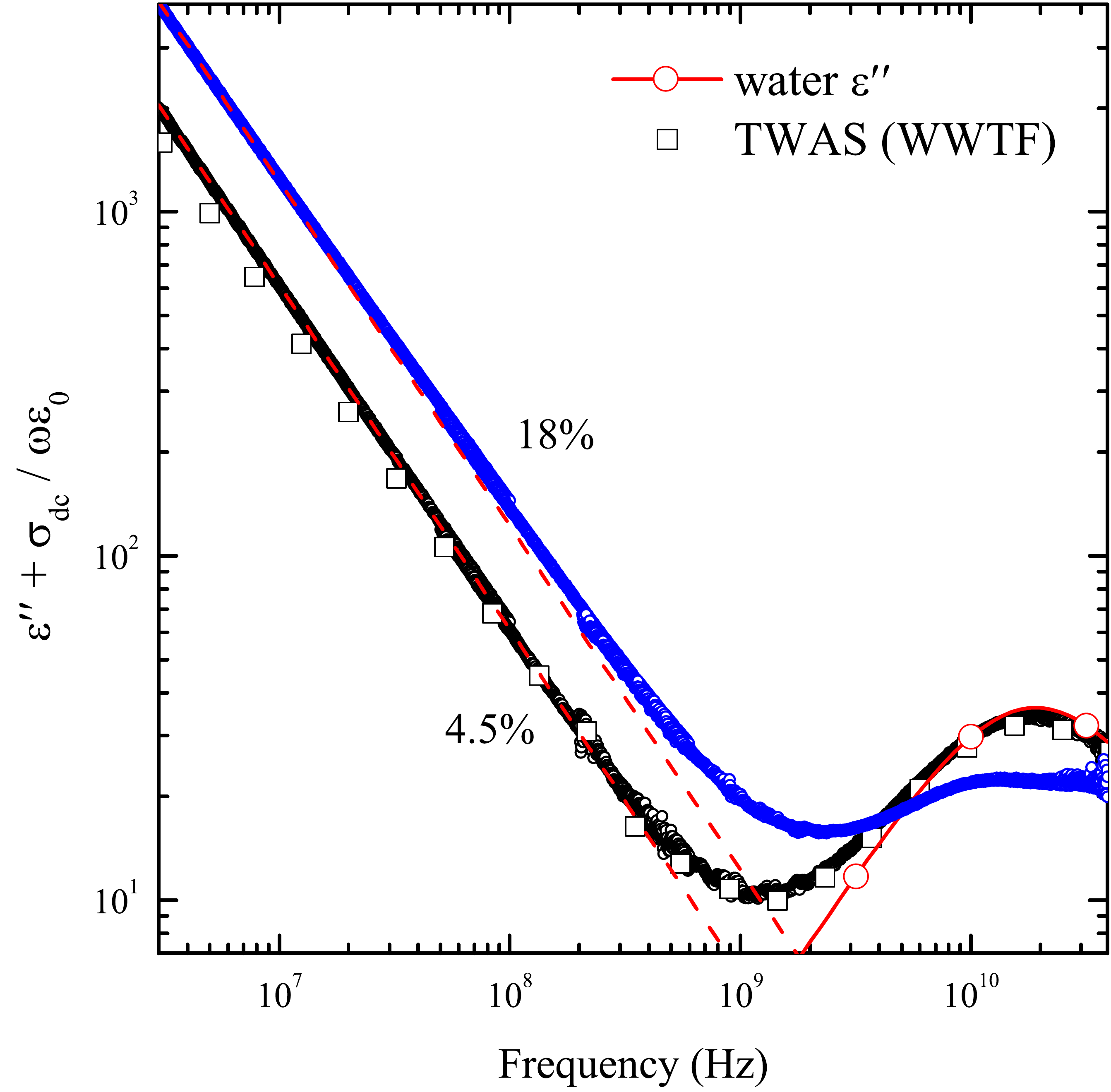}}
\caption{Plot of $\varepsilon^{\prime\!\prime}+\sigma_\mathrm{dc}/\omega\varepsilon_0$ of WAS as a function of frequency at 25$^\circ$C.
Electrode polarization corrections have negligible effect on these data.  The dashed lines have slope -1 and represent $\sigma_\mathrm{dc}/\omega\varepsilon_0$ contributions.  At 4.5\%, results for both sludge cake diluted with distilled water (black points) and TWAS from the WWTF (open squares) are shown.  The solid curve is $\varepsilon^{\prime\!\prime}$ of pure water.}
\label{fig:WASpermIm}
\end{figure}

Focusing now on Fig.~\ref{fig:WASpermIm}, $\varepsilon^{\prime\!\prime}+\sigma_\mathrm{dc}/\omega\varepsilon_0$ of both samples is dominated by the conductivity term below 1~GHz.
  The dashed lines in the figure follow a $\omega^{-1}$ frequency dependence and are drawn using $\sigma_\mathrm{dc}=0.34~\Omega^{-1}\mathrm{m}^{-1}$ and $\sigma_\mathrm{dc}=0.68~\Omega^{-1}\mathrm{m}^{-1}$ for the 4.5\% and 18\% samples respectively.  For frequencies below 100~MHz, linear fits to $\log\left(\varepsilon^{\prime\!\prime}+\sigma_\mathrm{dc}/\omega\varepsilon_0\right)$ as a function of $\log\omega$ were performed to extract $q$ from the observed $\omega^{-q}$ behavior.  For the 4.5\% sample, $q$ was $0.9961\pm 0.003$.  A value so close to one implies that $\varepsilon_\mathrm{r}$ exhibits very little dispersion over this frequency range, consistent with the observed $\varepsilon^\prime$ behavior. For the 18\% sample, on the other hand, with a value of $0.9728\pm 0.0003$, $q$ was noticeably less than one.  The only plausible explanation being that, in the higher concentration sample, dispersive mechanisms are active over this frequency range.  Above a few gigahertz, the conductivity term is negligible and the data are dominated by $\varepsilon^{\prime\!\prime}$.  As expected, the 4.5\% sample tracks the $\gamma$-dispersion of pure water very closely, whereas the 18\% sample shows significant deviations due to an active $\delta$-dispersion.  Again, the sludge cake sample that has been diluted with distilled water mimics the behavior of the TWAS sample from the WWTF.

\section{Summary \& Discussion}\label{sec:conclusions}

To the best of our knowledge, the electrical properties of WAS have not been previously reported.  We have preformed this characterization at two different concentrations of solid content over a broad range of frequencies and have identified the mechanisms leading to the observed dispersions.  The experimental and theoretical insights gained can be used to guide optimizations of the electromagnetic pretreatment of WAS.

Enhancements of the methane production rate are most likely to arise from punctured or otherwise compromised cell membranes that increase the bioavailability of the WAS to methane-forming microorganisms present during digestion.  At very high frequencies (tens of gigahertz) the WAS samples absorb energy from the electromagnetic fields via rotations of unbound dipolar water molecules.  This energy, dissipated as heat, will diffuse throughout the rest of the sample by conduction.  It is difficult to imagine how this microwave treatment would differ from a thermal cycling of WAS achieved, for example, using a hotplate.

The $\delta$-dispersion observed in the 18\% sample will be sensitive to electromagnetic treatments, between 100~MHz and 10~GHz.  Relaxation of protein-bound water molecules attached to cell membranes could be effective at initially concentrating the dissipated energy at the sites of the cells and could potentially induce cell lysis.  The reported benefits seen in anaerobic digestion when using WAS pretreated with electromagnetic irradiation at 915~MHz and 2.54~GHz may be in part a consequence of this effect.  Radio frequency treatments between hundreds of kilohertz and tens of megahertz are particularly intriguing as they will activate the Maxwell-Wagner effect causing the cell membranes to be repeatedly charged and discharged.  This process will also concentrate the initial dissipated energy at the cell sites and may be more effective at inducing cell lysis as the observed $\beta$-dispersion is significantly stronger than the $\delta$-dispersion.  Due to the higher concentration of suspended cells, $\beta$-dispersion is enhanced in the 18\% sample.  This observation suggests that electromagnetic pretreatments at radio frequencies are likely to be most effective when applied to samples having a high solid content.

As an outcome of this study, we are currently designing and building a series of custom high-power electromagnetic applicators that span a broad range of frequencies.  For each applicator, our permittivity and conductivity measurements allow matching circuits to be designed so as to achieve maximum power transfer from the source to the WAS load.  The overall efficiency of any practical large-scale applicator will be of critical importance.  The effectiveness of each applicator will be evaluated by quantifying cell lysis that results from the electromagnetic pretreatments and by measuring enhancements of the methane production rates from pretreated WAS during anaerobic digestion.

\ack

The financial support of the Natural Science and Engineering Research Council of
Canada Strategic Project Grant (\#396519-10) is gratefully acknowledged.  The Agilent 8722ES VNA used in this work was provided by CMC Microsystems and their support is gratefully acknowledged.

\end{document}